\documentclass{aastex}
\usepackage{spr-astr-addons}
\usepackage{url}

\usepackage{ulem}

\usepackage{cases}

\RequirePackage{color}

\newcommand{\dd}{\mathrm{d}}
\newcommand{\eq}[1]{eq. (\ref{#1})}

\newcommand{\nderrow}[2]{{\dd #1}/{\dd #2}}

\begin{document}

\title{Globular Cluster formation in a collapsing supershell}
\slugcomment{}
\shorttitle{GC formation in a collapsing supershell}
\shortauthors{Recchi et al.}

\author{S. Recchi\altaffilmark{1}} \and \author{R. W\"unsch\altaffilmark{1}} \and \author{J. Palou\v{s}\altaffilmark{1}} 
\and \author{F. Dinnbier\altaffilmark{1}}
\affil{Astronomical Institute, Academy of Sciences of the Czech Republic, Bo\v{c}n\'{i} II 1401-2a, Prague, Czech Republic\\ Email: simone.recchi@asu.cas.cz}

\begin{abstract}
  Primordial clouds are supposed to host the so-called population III
  stars. These stars are very massive and completely metal-free. The
  final stage of the life of population III stars with masses between
  130 and 260 solar masses is a very energetic hypernova explosion.  A
  hypernova drives a shock, behind which a spherically symmetric very
  dense supershell forms, which might become gravitationally unstable,
  fragment, and form stars. In this paper we study under what
  conditions can an expanding supershell become gravitationally
  unstable and how the feedback of these supershell stars (SSSs)
  affects its surroundings. We simulate, by means of a 1-D Eulerian
  hydrocode, the early evolution of the primordial cloud after the
  hypernova explosion, the formation of SSSs, and the following
  evolution, once the SSSs start to release energy and heavy elements
  into the interstellar medium. Our results indicate that a shell,
  enriched with nucleosynthetic products from SSSs, propagates
  inwards, towards the center of the primordial cloud. In a time span
  of a few Myr, this inward-propagating shell reaches a distance of
  only a few parsec away from the center of the primordial cloud. Its
  density is extremely high and its temperature very low, thus the
  conditions for a new episode of star formation are achieved. We
  study what fraction of these two distinct populations of stars can
  remain bound and survive until the present day. We study also under
  what conditions can this process repeat and form multiple stellar
  populations. We extensively discuss whether the proposed scenario
  can help to explain some open questions of the formation mechanism
  of globular clusters.
\end{abstract}


\section{Introduction}
The evolution of a spherically symmetric blast wave expanding into the
interstellar medium (ISM) is a well-established, textbook topic, amply
covered in many highly detailed reviews \cite[see e.g.][]{omk88,
  bks95}.  After the relatively short free expansion and adiabatic
phases, the gas just behind the outward-propagating shock cools and
forms a very dense shell of material which continues propagating
outwards due to its inertia.

The stability of this thin shell of gas has been also extensively
studied \cite[e.g.][see also Sect. \ref{sec:review}]{vis83, whit94a,
  whit94b, elme94, ehle97, wp01, franta17}. Unstable modes can be
found for a wide range of initial conditions, and the fragmentation of
the unstable supershell can lead to star formation.  This has been
proposed as a possible triggering mechanism for star formation
\cite[e.g.][]{gs78, ee78, palo94}. Stars formed in this still
outward-propagating supershell will evolve and release winds and,
possibly, supernova (SN) explosion, into the ISM like any other
stellar system. However, the evolution of the supershell and of the
stellar ejecta after the formation of the supershell stars (SSSs) is
still largely unexplored. Our intention is to fill this gap.

Gravitationally unstable supershells may also be related to the
formation of globular clusters (GCs). In particular, the lack of GCs
with a metallicity below [Fe/H]$\simeq$ -2.5 \citep{harr96} has
puzzled astronomers for a long time. It seems possible to explain this
fact by assuming an early generation of stars at the center of a
proto-GC. Supernovae go off and create an expanding supershell.  The
mixing of the expanding supershell with the ejecta of the first
generation of stars lead to the right metal enrichment \citep{bbt91,
  parm99, rd05, sw00}. In particular, if the mass of the proto-GC is
larger than some threshold (usually in the range of 10$^6$-10$^7$
M$_\odot$), the fragmentation of the supershell can occur before the
supershell becomes gravitationally unbound. This guarantees that the
newly-formed SSSs will remain bound to the cluster.

However, it is unlikely that this scenario can explain all chemical
properties of GCs. These properties are reviewed in Sect.
\ref{sec:review}, but in brief GCs are characterized by multiple
stellar populations and by characteristic anti-correlations between
some pairs of light elements. Particularly remarkable and ubiquitous
are the Na-O and Mg-Al anticorrelations \cite[see e.g.][and references
therein,]{sned97, gsc04} notice however that the Mg-Al anticorrelation
is ubiquitous only in massive GCs and often is missing in less massive
ones \cite[see e.g.][and references therein]{panc17}. On the other
hand, there is a remarkable homogeneity in the abundances of iron-peak
elements in the large majority of observed GCs. The scenario depicted
above seems thus unlikely because it implies that the stars which
created and enriched the expanding supershell in iron either
disappeared or had already the relatively high [Fe/H] we observe
nowadays. Other scenarios have therefore been proposed to explain the
peculiar chemical properties of GCs.  These scenarios are very briefly
reviewed in Sect. \ref{sec:review}.

However, it is maybe possible to save the broad scenario
(fragmentation in the expanding supershell) by invoking a different
trigger for the supershell expansion, namely a primordial, very
energetic population-III (Pop-III) star. This can also easily explain
the relatively high iron content of observed GCs \citep{beas03}. The
stars formed in this expanding supershell will evolve as any normal
star: they will create stellar winds and, depending on the IMF, some
of them will explode as Type II supernovae (SNeII). Some of the winds
of these SSSs will propagate inwards rather than outwards. They might
accumulate in the center (or close to the center) of the proto-GC,
cool and condense there, maybe leading to a new episode of star
formation. This is the general idea we want to explore here and in
follow-up papers. The paper is structured as follows: in Sect.
\ref{sec:review} we will briefly summarize the main physical and
chemical properties of GCs and what features a theory of GC formation
must have in order to reproduce them. We will then lay down the idea
of the paper in more detail in Sect. \ref{sec:idea}, linking the
proposed model to the requisites of theories of GC formation described
in Sect. \ref{sec:review}.  The initial conditions are presented in
Sect. \ref{subs:incon}, the adopted numerical scheme in Sect.
\ref{subs:numer}, the assumed criteria for the fragmentation of the
expanding supershell in Sect. \ref{subs:fragm}, and the results of
numerical experiments in Sect. \ref{sec:results}. Specifically, in
Sect.~\ref{sub:expss} we study the expanding supershell and the
formation of the 1G SSSs, in Sect. \ref{sub:inwss} we study the fate
of the inward-propagating supershell formed by the SSSs winds, and we
establish under what conditions this supershell can form a new
generation of stars and what are the characteristics of these
second-generation stars. Finally, in Sect. \ref{sec:disc} we will
discuss our results and draw conclusions.

\section{The properties of globular clusters}
\label{sec:review}
GCs show characteristics which have puzzled astronomers for decades.
Many theories of their formation have been put forward but none of
them manage to explain all properties of GCs. This lead \cite{bast15}
to write "Hence, with the exclusion of all current models, new
scenarios are desperately needed."  Many review papers describe in
detail the chemical characteristics of GCs and the properties a theory
of GC formation should have in order to conform with observations
\cite[see in particular][]{bast15, renz15}. We will follow in
particular \citet{renz15}, grossly simplifying the description. We
refer to the original paper for more detail.

\begin{enumerate}
\item {\bf Specificity}. A crucial property of GCs is the presence of
  multiple populations of stars, with a number of different stellar
  populations ranging from two to perhaps seven \citep{milo15}. This
  phenomenon appears to be specific of GCs: young massive clusters do
  not show clear evidence of multiple populations although the masses
  of these clusters are similar \cite[see recent papers by][]{bast13a,
    cabr16}. It seems thus, to quote \citet{renz15}, that "special
  conditions encountered only in the early Universe appear to be
  instrumental for the occurrence of the GC multiple population
  phenomenon." {\it We identify in this work these "special
    conditions" with the explosion of very massive pop-III stars}, as
  we describe in detail in Sect. \ref{sec:idea}.  Another aspect of
  the specificity of GCs is the fact that field halo stars with the
  same metallicity and age do not show the same chemical
  characteristics of GC stars. On the other hand, we note that
  \citet{niederhofer17} recently reported the discovery of multiple
  stellar populations in 3 intermediate-age star clusters in the Small
  Magellanic Cloud.

\item {\bf Predominance}. At least half of the stars in GCs have been
  self-polluted within the GC, thus do not belong to the first
  generation (1G) of stars. This is strictly connected to the
  so-called mass-budget problem: a significant mass in ejecta from 1G
  pollutant stars is required to explain the fraction of
  late-generation (LG) stars but, for ordinary IMFs, 1G ejecta
  constitute only a small fraction of the total 1G mass. The fraction
  of 1G stars could be as low as one third and pretty constant among
  different GCs \citep{bl15}, although a more recent study shows a
  clear variability of this fraction and an anti-correlation with the
  GC mass \citep{milo17}. In particular, the smallest GCs show 1G
  fractions larger than 50\%, whereas this fraction decreases to about
  20\% for the most massive GCs. Notice that the mass dependence is
  not limited to the 1G stellar fraction. Also the spread in helium
  \citep{milone15} and in light elements \citep{milo17} seems to
  correlate with the GC mass.

\item {\bf Discreteness}. The process of star formation in GCs appears
  to be discontinuous, characterized by well-separated events. This is
  clearly visible in well-separated main sequences in color-magnitude
  diagrams \citep{piot07}. This is much less clear in the apparently
  continuous Na-O and Mg-Al anticorrelations, although at least in the
  Na/O plane discrete multiple populations are discernible
  \citep{mari08, milo15b} and signs of patchiness in the distribution
  of other chemical elements have started to emerge \citep{carr15}.

\item {\bf Supernova avoidance}. As already mentioned, stars in the
  large majority of GCs do not show a spread in iron-peak or in other
  heavy elements, clearly showing that SN ejecta should negligibly
  contribute to the chemical enrichment of LG stars. However, it is
  worth mentioning that the fraction of observed GCs which shows also
  a spread in heavy elements is increasing by the day and nowadays
  amounts to about 15 to 20\% of the observed GCs \cite[see in
  particular][and references therein]{mari15}.  This variation is not
  limited to iron-peak elements but it is extended to calcium and
  s-process elements. This has been shown in a number of objects
  \cite[see e.g.][]{mari11, john17, milo15, yong14, yong16}. SNe might
  therefore play after all a role in the evolution of some GCs.
\end{enumerate}

There are other peculiar characteristics of GC stars and constraints
for theories of GC formation. In particular, a theory of GC formation
should be able to reproduce the above mentioned Na-O and Mg-Al
anticorrelations, and should reproduce the observed spread in He. We
will not treat these aspects in detail, as they are more closely
related to the nucleosynthesis of stars within a GC rather than to the
mechanism of GC formation.

As already mentioned, various models have attempted to explain the
peculiarities of GC stars. We refer again to review papers for a
detailed description. We recall here very briefly only the two more
widely investigated scenarios:
\begin{itemize}
\item {\bf Pollution from winds of massive stars, e.g. from fast
    rotating massive stars (FRMS)} \citep{cc16, krau13, decressin07,
    wuen16, richa16, lt17}. The idea is that extruding disks of FRMS,
  filled with ejecta from stellar winds, mix with pristine gas and
  form new generations of stars.
\item {\bf Pollution from AGB stars} \citep{derc08, derc10} Here the
  assumption is that the ejecta of AGB stars, given their low
  velocity, might remain bound to the proto-GC, whereas SN remnants do
  not. The AGB ejecta, possibly mixed with some pristine material,
  form LG stars.
\end{itemize}
Both these scenarios, and many other scenarios not described here
\cite[see e.g.][]{demi09, bast13b, deni15} have attractive features,
as well as some problems which require some tweaking of parameters.

\section{The proposed model}

\subsection{General idea}
\label{sec:idea}

\begin{figure*}[t]
\includegraphics[width=16cm]{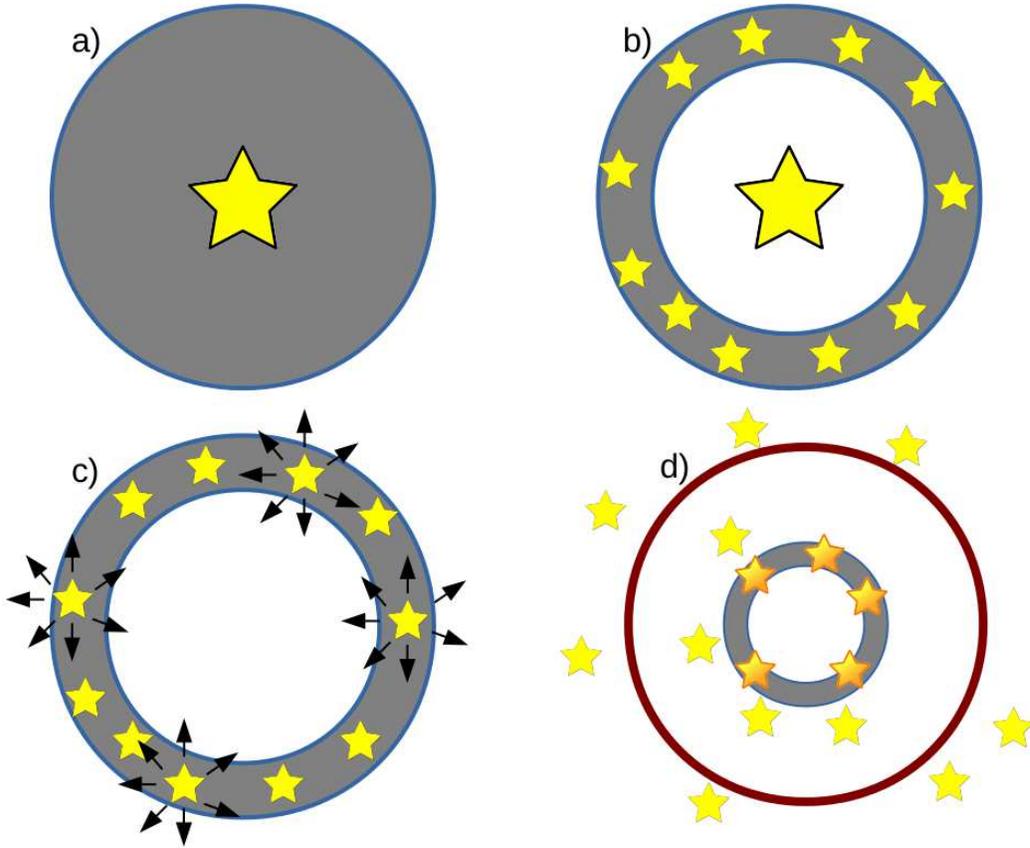}
\caption{The major phases of our assumed scenario for the formation of
  GC: {\it a)} A very massive pop-III star explodes in the middle of a
  primordial cloud, {\it b)} The energy of this star produces a shell
  of dense, cold gas. This supershell becomes gravitationally
  unstable, fragments and forms new stars (1G stars) {\it c)}
  Supershell stars release energy, due to stellar winds and SN
  explosions. Some of the gas they release is driven inwards.  {\it
    d)} This triggers the formation of a new, inward- propagating
  supershell, which fragments and forms a second generation of stars.
  The thick red line denotes the tidal radius in the external
  proto-galactic field.  }
\label{fig:sketch}
\end{figure*}

We recall here the main aim of our paper. We want to study the
fragmentation of a supershell created by the explosion of a Pop-III
star and the fate of winds of stars formed out of this fragmented
supershell. It is to be expected that approximately a half of these
winds will propagate towards the center of the system (the place where
the Pop-III star exploded), perhaps creating the conditions for a new
episode of star formation. In fact, the density of gas piling up close
to the center must increase, thus the cooling increases too, leading
to large regions of cold, dense gas, probably prone to the formation
of new stars. Alternatively, the inward-propagating winds of SSSs
could trigger the formation of another inward propagating dense
supershell, able to fragment itself and form a new generation of
stars. At the same time, the SSSs destroy by their winds the primary
expanding shell and a fraction of them leaves the proto-cluster due to
their outward velocity inherited from the expanding shell, in
combination with galactic tides. This process might then repeat
itself. We illustrate the main phases of our assumed scenario in Fig.
\ref{fig:sketch}.

We believe that this study is interesting per se, as we are not aware
of similar studies in the literature, but we want to explore here the
possibility that this mechanism is connected to the formation of GCs.
We will thus go through the points illustrated in Sect.
\ref{sec:review} (specificity, predominance, discreteness, SN
avoidance) and analyze if our proposed scenario might help explaining
them. Ours is thus not properly a scenario of progenitors of polluted
GC stars (as the massive stars and the AGB scenarios are). We think
that many authors (many of them mentioned above) have made a great job
trying to identify the nucleosynthesis pathways towards the chemical
patterns observed in GCs. We simply devise a possible evolutionary
pathway which might help solve some of the difficulties encountered by
these progenitor models. For convenience we tune our discussion and
the following calculations to the massive stars scenario, but we
believe that with simple changes our idea can be adapted to the AGB
scenario (and perhaps to the other proposed scenarios) as well.

\begin{enumerate}
\item {\bf Specificity}. This is easy: stars in GCs appear unlike
  stars in young massive clusters because nowadays Pop-III stars do
  not exist any more, thus our proposal can apply preferentially to
  very old massive clusters, not to young ones. Our scenario requires
  a very energetic Pop-III star, able to sweep up a large mass of gas.
  The properties and nucleosynthesis of these extremely massive
  (M$>$100 M$_\odot$) stars are pretty well known \cite[e.g.][]{hw02}.
  Attention in the last few years has shifted to less massive (M$<$100
  M$_\odot$) first stars \cite[e.g.][]{hw10, chop16}. This is most
  probably due to the need to explain the observed chemical
  composition of ultra metal-poor stars \cite[e.g.][]{plac16}, which
  are extremely poor in iron, thus can not have been polluted by very
  massive Pop-III stars. It is nevertheless still true that, due to
  the lack of metals and to the higher temperature of the cosmic
  microwave background, the conditions in the early universe should
  have favoured the formation of very massive stars \citep{bromm99}.
  Therefore it appears to us plausible that very massive Pop-III stars
  existed in the early universe, and were responsible for the
  formation of GCs. The high iron yields of theses Pop-III stars are
  responsible for the relatively high [Fe/H] in observed GCs
  \citep{beas03}. Less massive (and less energetic) Pop-III stars
  were, on the other hand, unable to sweep up large masses of pristine
  gas and were thus unable to trigger the formation of large clusters.
  They polluted the ISM more patchily, creating the conditions for the
  formation of ultra metal-poor stars. Recently, \cite{elme17}
  suggested that the different conditions that allowed the formation
  of GCs in the early universe were due to the higher densities. This
  is of course a viable suggestion, alternative to our proposed
  Pop-III triggered mechanism.

\item {\bf Predominance}. The problem of the predominance (and the
  associated problem of the mass budget) is typically solved assuming
  that the 1G stars we see nowadays in a GC are just a small fraction
  of the polluting stars originally present in the proto-cluster.
  However, no convincing scenario has been put forward to explain why
  a large fraction of 1G stars should leave the proto-GC. Our proposed
  scenario might provide a partial solution because $(i)$ the SSSs
  possess a positive radial velocity (they inherit the velocity of the
  supershell). Some of the SSSs might have a velocity larger than the
  escape velocity from the proto-GCs. On the other hand, the winds
  from these SSSs will have at least partially a negative radial
  velocity, assuming that they expand isotropically from the SSSs and
  that their velocities are larger than the stellar velocities. $(ii)$
  Even if SSSs do not escape directly, they are formed at relatively
  large distances from the center of the cluster.  As they start being
  pulled towards the cluster center, they will acquire relatively
  large velocities. The velocity dispersion of SSSs (the 1G stars in
  our scenario) will be thus larger than the velocity dispersion of
  stars formed closer to the center of the cluster, thus it is less
  likely that they will be bound to the cluster. At least in
  principle, our model can explain the anticorrelation between
  cluster mass and 1G star fraction (and between mass and chemical
  complexity) described in Sect. \ref{sec:review}.  According to our
  model, a crucial parameter is the fragmentation time of the
  supershell (see Sect. \ref{subs:fragm}); for large fragmentation
  times, the supershell can propagate further and collect more mass.
  The larger mass in SSSs does not necessarily imply a larger mass in
  1G stars. Having formed at large radii, a large fraction of the SSSs
  might be unbound and leave the proto-GC. On the other hand, the SSS
  ejecta are more massive and therefore a larger fraction of LG stars
  can be formed. This can explain the anti-correlation between cluster
  mass and 1G star fraction.  Moreover, as already mentioned (and as
  we will see below), a new generation of stars can be formed in an
  inward-propagating shell.  The radius at which this shell will
  fragment and form new stars will depend on the radius at which SSSs
  form. If the fragmentation radius of the inward-propagating shell is
  large enough, this process can repeat itself (see also below).

\item {\bf Discreteness}. It is obvious that 1G stars and LG stars are
  well-separated. 1G stars are the surviving SSSs, whereas LG stars
  are formed later, closer to the center of the cluster. To explain
  the presence of multiple ($n>2$) separated generations of stars, we
  must assume that the mechanism we describe here might replicate
  itself: SSSs form an inward-propagating shell, which fragments and
  form new stars, which create another inward-propagating supershell,
  and so on. As we will see in Sect. \ref{sub:inwss}, simulations
  suggest that this mechanism might be possible. Moreover, as
  explained above, this mechanism might depend on the fragmentation
  time of the supershell (and thus on the radius at which SSSs form).
  This can affect the fragmentation radius of the inward-propagating
  shell, which in turn will affect the probability that this mechanism
  repeats itself. We see thus, at least in principle, a possible
  explanation for the correlation between GC mass and chemical
  complexity (spread in helium and in light elements) described in
  Sect. \ref{sec:review}. The larger the fragmentation time, the
  larger the collected supershell mass, the larger the mass in LG
  stars, the larger the probability of forming multiple (and not just
  two) stellar population, the larger the chemical complexity.

\item {\bf Supernova avoidance}. In the framework of the massive
  stars, we must assume that the processes leading to the formation of
  LG stars should occur relatively quickly (within few Myr), in order
  to be completed before SNeII go off. As we will see below, again,
  simulations (and analytical considerations) seem to support this
  view. However, as remarked in Sect. \ref{sec:review}, a
  non-negligible fraction of GCs do show a spread in heavy elements,
  which may be due to differences in their supplies by Pop-III stars
  of different mass, or due to changing fragmentation time of the
  supershell.  Some of these GCs (as originally suggested for the
  archetypal examples of Omega Centauri and M22) might be remnants of
  tidally disrupted dwarf galaxies \cite[see e.g.][]{mari15} but, for
  some others, an internal mechanism might be invoked.  This is easily
  accounted for, if the formation of LG stars lasts more than a few
  Myr (see also Sect.  \ref{sub:inwss}).

\end{enumerate}

After having laid down the general idea of this paper, we describe
here and in the following section the calculations we have performed
to substantiate it. The calculations presented here should be seen as
a feasibility study rather than as a complete, self-consistent
description of the phenomena we wish to simulate. A fully consistent
simulation requires the inclusion of many complex physical processes
and is very computationally demanding. Before embarking on this kind
of computation, we want to be sure that the proposed scenario can
really work, thus we will present here a simplified model. Even this
simplified model turns out to be more computationally demanding than
we thought. We will present more detailed simulations in forthcoming
papers.

\subsection{The initial conditions}
\label{subs:incon}
The parameters of the Pop-III stars are taken from \cite{hw02}. We
concentrate on the model having a He core mass of 130 M$_\odot$,
corresponding to an initial mass of about 260 M$_\odot$. The explosion
energy of this model Pop-III star is about 10$^{53}$ erg. This model
Pop-III star releases 40 M$_\odot$ of Fe. If mixed with 10$^7$
M$_\odot$ of pristine gas, this amount of iron leads to a [Fe/H] of
about -2.5, which is the minimum [Fe/H] observed in GCs. Lacking a
detailed description of mixing processes, we will assume henceforth
that the Pop-III explosion leads instantaneously to a uniform
metallicity of [Fe/H]=-2.5 in the supershell, which is thus also the
metallicity of our 1G stars. As we will see below, the mass of gas
accumulated in the supershell is generally less than 10$^7$ M$_\odot$,
thus the supershell metallicity might be higher than -2.5. Our choice
is thus conservative: cooling and star formation might be more
efficient than we assume in what follows.

Several papers in the literature have been devoted to the study of the
formation and explosion of very massive Pop-III stars \cite[see
e.g.][]{abel02, ripa02, op03, bromm09, humm16}. We choose here to fix
the initial conditions of the ISM surrounding the Pop-III star sitting
in the middle of an ISM cloud according to the results of
\citet{yosh06}. In their study, the ISM clouds are characterized by a
very small core, a very high central density and a $r^{-2}$ envelope
beyond the core. Namely the number density is characterized by the
following profile:

\begin{equation}
n(r)=\begin{cases}n_{\rm c}\;\;\;\;\;{\rm if }\;r<r_{\rm c}\\ n_{\rm c}\left(\frac{r}{r_{\rm c}}\right)^{-2}
\;\;\;\;\;{\rm if }\;r\geq r_{\rm c}\end{cases}
\label{eq:profile}
\end{equation}
\noindent
Typical values found in the simulations of \citet{yosh06} are $r_{\rm
  c}=1$ au for the core radius and $n_{\rm c}=2\cdot 10^{15}$
cm$^{-3}$ for the central number density. This central density is very
high, but it falls off very rapidly. The total mass contained in the
whole computational grid (a sphere of radius 200 pc) is $\sim
3\cdot 10^6$ M$_\odot$. Therefore, by adopting these initial
conditions we are unable with the mass available to reproduce the most
massive GCs observed in the Galaxy, i.e. those with masses of the
order of 10$^6$ M$_\odot$.  We consider a minimum temperature of the
ISM of 50 K. If dictated by the cosmic microwave background alone,
this temperature implies a redshift of formation of the Pop-III star
of about 17.

\subsection{The numerical code}
\label{subs:numer}
We construct a 1-D, Eulerian, spherically symmetric code to follow the
expansion and fragmentation of the supershell produced by the Pop-III
explosion. Of course, it is highly unlikely that the system will
maintain a perfect spherical symmetry during the whole evolution.  As
mentioned before, we plan here to perform a simplified feasibility
study; more detailed 3-D simulations are planned and will be presented
in forthcoming papers. The code, which is second-order accurate, is
based on a HLLC Riemann solver and uses slope limiters to avoid
spurious oscillations \citep{toro09}. The cells are uniform and 0.01
pc wide.  This means that the core of 1 au in the initial ISM
distribution is not resolved in our simulation. As already mentioned,
we avoid the (uncertain) simulation of the process of chemical
enrichment of the supershell due to the Pop-III ejecta and consider
from the beginning a constant metallicity of [Fe/H]=-2.5. The physical
processes included in our simulations are:
\begin{itemize}
\item Molecular and atomic cooling. For T $>$ 10$^{3.8}$ K we adopt
  the cooling function and the ionization fractions of \citet{schu09}
  with fitting formulae taken from \citet{voro15}.  Below 10$^{3.8}$ K
  we adopt the atomic and molecular cooling of \citet{dm72}. The
  ionization fractions are adapted from \citet{abel97}.
\item Self-gravity. The gravitational acceleration is calculated as
  ${g}=-{GM(r)}/{r^2}$, where G is the gravitational constant and
  $M(r)$ is the mass in gas and stars contained within radius $r$.
\item Thermal conduction. A Spitzer-type thermal conduction equation
  is solved by means of the Crank-Nicolson method. We refer to
  \citet{bd86} for the numerical implementation of this method.
\end{itemize}
We follow the expansion of a supershell created by the Pop-III star
until it fragments and forms 1G stars. Later, we follow the evolution
of the system when SSSs start inject energy and chemical elements into
the ISM. Thus we include in the simulations two additional physical
processes:
\begin{itemize}
\item Feedback from a single stellar population. This is calculated by
  means of a tailored Starburst99 simulation \citep{leit99, leit14}.
  The IMF of the SSS is assumed to be the Kroupa IMF, as it closely
  resembles the spectrum of mass fragments in an unstable supershell
  \citep{tt03}. The energy produced by SSSs is inserted as thermal
  energy. Energy, gas mass, and mass in chemical elements produced by
  the SSSs are inserted in those grid cells, where stars reside (see
  below for the dynamics of the stars). The amount of feedback to be
  inserted in each cell is proportional to the total stellar mass in
  any given cell.
\item Dynamics of stars. Here, we will simply assume that the stars
  inherit the velocity of the supershell at the moment of
  fragmentation. Subsequently, their dynamics is dictated by the
  competition between their inertia and the gravitational pull. The
  stellar population formed in the supershell is subdivided into 250
  mass bins, each of which evolves independently. To guarantee a
  spread in the stellar bins, we distribute the velocities according
  to a Maxwellian distribution, centered on the supershell expansion
  velocity.  Of course, our treatment grossly simplifies the dynamics
  of the stars, and only detailed N-body simulations can shed light on
  the final fate of SSSs.
\end{itemize}
Many other physical phenomena which might be relevant for the problem
at hand have not been considered. As already mentioned, this has been
done on purpose, in order to simplify the treatment and put more
emphasis on the feasibility of the proposed scenario. A critical
assessment of the missing physics will be presented in Sect.
\ref{sec:disc}.

\subsection{The fragmentation of the supershell}

\label{subs:fragm}

The ambient ISM is unlikely to be perfectly isotropic, and when
accreted, it seeds the supershell with perturbations in its surface
density. We assume that the amplitudes of perturbations are small
leaving the spherically symmetric model of the expanding supershell
still justified. Since our one dimensional model cannot follow shell
fragmentation in three dimensional space, we estimate the time
$t_{frg}$ of the supershell fragmentation as an instant when the shell
expansion time $t$ equals to the collapse time-scale $t_{col}$ of the
most unstable fragment in the shell.  Note that the fragment collapses
in the direction tangent to the shell surface.  Following
\citep{whit94a}, the collapse time-scale is
%
\begin{equation}
t_{col} = \left\{ \frac{G \Sigma}{d} - \frac{a_s^2}{d^2} \pm \frac{\alpha^2}{t^2}  \right\}^{-1/2},
\label{efrag}
\end{equation}
where $\Sigma$ is the surface density, $d$ the fragment radius, $a_s$
the sound speed inside of the shell, and $\alpha$ the exponent in the
shell expansion law,
\begin{equation}
R = K t^{\alpha}.
\label{eexpansionlaw}
\end{equation}
The first term on the right hand side of \eq{efrag} expresses the
tendency of the self--gravity to bind the fragment together while the
second is due to the thermal pressure gradient acting in the opposite
direction. The third term describes either stretching of the fragment
in the case of the expanding shell (negative sign), or the fragment
contraction in the case of the collapsing shell (positive sign). The
latter case applies to the secondary inward propagating shell formed
by the 1G SSSs.

In the case of a supernova explosion at the density peak of a medium
with a power--law density profile following equation
(\ref{eq:profile}), the exponent of the expanding law in the energy
conserving phase is $\alpha$ = 2/3 and in the momentum conserving
phase $\alpha = 1/2$ \citep{omk88}. The high density in our model
results in rapid cooling, so the supershell switches from the energy
conserving phase to momentum conserving phase very early (see also
\citet{haid16}), and fragmentation takes place in the momentum
conserving phase with $\alpha = 1/2$.

The relative importance of the self--gravity and external pressure in
confining the supershell significantly influences the fragmentation
process and the supershell fragmentation time $t_{frg}$
\citep{franta17}. In Appendix \ref{apppressure}, we show that the
supershell likely fragments when dominated with its self--gravity, in
which case the collapse time-scale of the most unstable fragment in
the supershell is
\begin{equation}
t_{col} = \left(\frac{G^2 \Sigma^2}{4 a_s^2} \pm \frac{\alpha^2}{t^2}\right)^{-1/2}.
\label{eq:fragtime}
\end{equation}
During the simulation, we evaluate $t_{col}$ at each time step, and
identify the fragmentation time $t_{frg}$ as the time when the
expansion time $t$ exceeds the collapse time-scale $t_{col}$ for the
first time. We assume that the SSSs are formed at this instant.  The
surface density, $\Sigma$, and the sound speed, $a_s$, in the shell
are measured in the simulation by averaging between the inner and the
outer edges of the shell, detected as the steepest positive and
negative density gradients.  Note that the sound speed is almost
always constant with a value corresponding to the temperature $50$\,K
as the shell is nearly isothermal (see Figure \ref{fig:ss}).

\section{Results}
\label{sec:results}

Here we present results of 1D simulations modelling the supershells
formed after an explosion of a Pop-III star releasing energy
$10^{53}$\,ergs into a centre of a gaseous cloud with the density
profile given by eq.~(\ref{eq:profile}). The primary expanding
supershell leading to the formation of 1G stars is described in
Sect.~\ref{sub:expss}. The secondary, inward propagating supershell
out of which the LG stars form is described in Sect.~\ref{sub:inwss}.
Two versions of the model are studied differing in the location within
the primary supershell where the 1G SSSs are inserted.

\subsection{The expanding supershell and the formation of 1G stars}
\label{sub:expss}

\begin{figure}[t]
\includegraphics[width=5.6cm, angle=270]{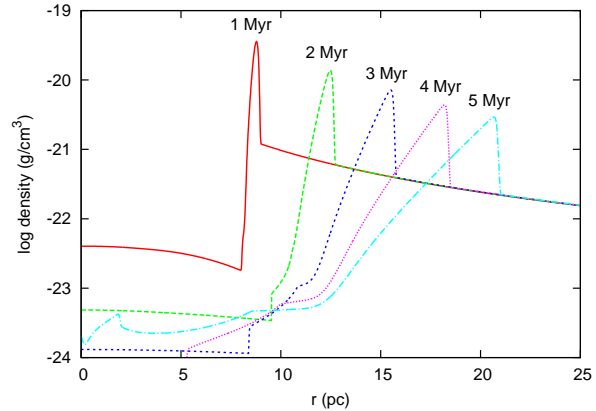}
\caption{Evolution of the gas density (in g cm$^{-3}$) after a Pop-III
  explosion. The density is plotted every Myr.}
\label{fig:dens_expl}
\end{figure}

\begin{figure}[t]
\includegraphics[width=5.6cm, angle=270]{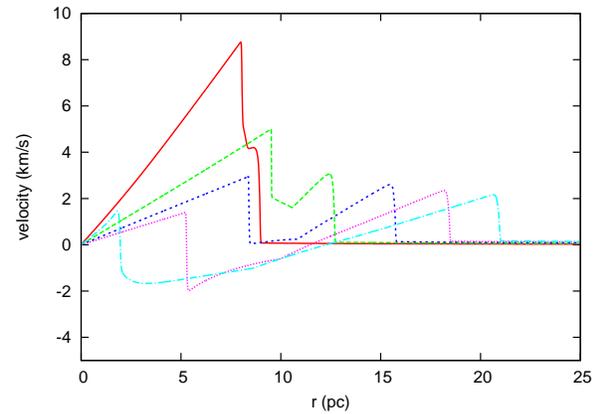}
\caption{Gas velocity (in km/s) after a Pop-III explosion, plotted
  every Myr as in Fig.~\ref{fig:dens_expl}.}
\label{fig:vel}
\end{figure}

\begin{figure}[t]
\includegraphics[width=5.6cm, angle=270]{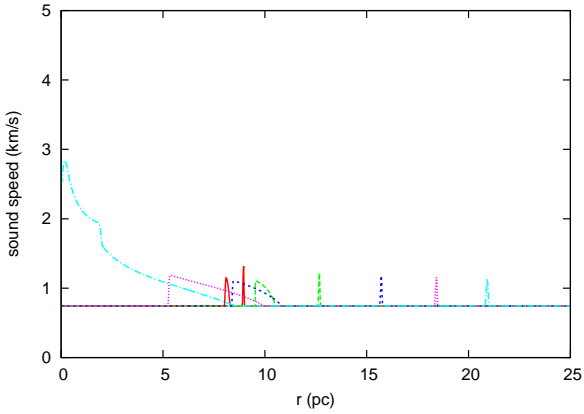}
\caption{Sound speed (in km/s) after a Pop-III explosion, plotted
  every Myr as in Fig.~\ref{fig:dens_expl}.}
\label{fig:ss}
\end{figure}

\begin{figure}[t]
\includegraphics[width=5.6cm, angle=270]{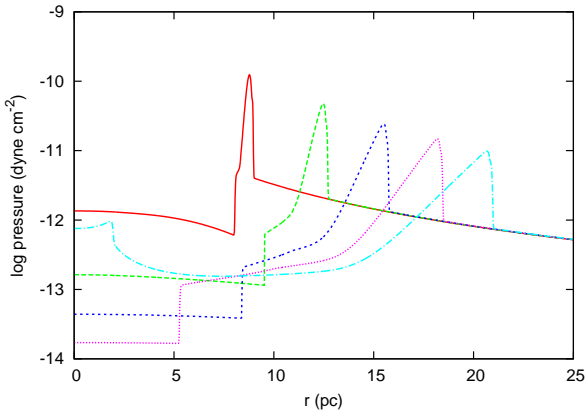}
\caption{Logarithm of the pressure (in cgs) after a Pop-III explosion,
  plotted every Myr as in Fig.~\ref{fig:dens_expl}.}
\label{fig:pres}
\end{figure}

The evolution of the ISM density following the Pop-III explosion is
shown in Fig. \ref{fig:dens_expl}. As mentioned in Sect.
\ref{subs:fragm}, we expect the external shock to expand approximately
as $t^{1/2}$.  A reverse shock starts to appear in the last
time-frames of this plot.  Figure \ref{fig:vel} shows the velocity,
which decreases with time as expected. Figure \ref{fig:ss} shows the
sound speed $a_{\rm s}$ for the same time frames as in the previous
plots. Apart from a very narrow spike, the sound speed within the
supershell is constant, because the efficient cooling leads rapidly to
the temperature equal to the minimum temperature of 50 K in the
shocked supershell gas. The constancy of the sound speed justifies the
considerations and the analytical estimates made in the previous
section. Fig. \ref{fig:pres} shows the pressure profile for the same
time frames.

In this model, the fragmentation of the supershell occurs, according
to the condition \eq{eq:fragtime}, at $t_{frg} \simeq$ 4.8 Myr, an
instant in which the external shock has reached the radius $\sim$ 20
pc.  This corresponds to the dotted-short dashed line in Figs.
\ref{fig:dens_expl} and \ref{fig:vel}.  At this instant, the
supershell has a mass of slightly more than $3 \cdot 10^5$ M$_\odot$.
Assuming a star formation efficiency of 30\%, we obtain a mass in SSSs
of $\sim$ 10$^5$ M$_\odot$. Since we expect the large majority of
these SSSs to become unbound (see the discussion in Sect.
\ref{sec:idea}), even considering LG stars we will not be able to form
a GC larger than a few times 10$^4$ M$_\odot$.  This is consistent
with the low-mass end of the mass distribution function of observed
GCs; it seems much harder to obtain masses of the order of $\sim$
10$^6$ M$_\odot$, corresponding to the most massive observed GCs.

We have also tried to run models with densities larger than the ones
described in Sect. \ref{subs:incon}, namely with a central density of
$n_{\rm c}=6\cdot 10^{15}$ cm$^{-3}$. In this case, the fragmentation
of the supershell occurs earlier, at $t_{frg} \simeq$ 0.6 Myr
(\eq{eq:fragtime}).  However, given the higher density, the mass
accumulated in the supershell at this moment in time is more or less
the same.  With some fine-tuning, we are able to increase the mass of
the supershell by a factor of a few, but with our assumptions and
initial conditions it seems impossible to accumulate substantially
more mass in the supershell. It seems thus that the problem of having
a supershell massive enough has more to do with the profile rather
than to the central density. We have tried different, flatter initial
ISM density distributions and we can produce in this way unstable
supershells with masses up to $\sim$ 10$^7$ M$_\odot$. Notice that the
$r^{-2}$ profile found by \citet{yosh06} extends only up to a few pc,
thus a different and perhaps flatter gas distribution outside the
first few pc is conceivable. We prefer in this work to stick to the
\citet{yosh06} initial conditions in order to study the basic scenario
in a simple and well-justified setting. We will change the initial
conditions in forthcoming papers. In the meantime we note that a
flatter initial gas distribution would not only allow a larger total
GC mass, but might also help to capture more SN ejecta, because it
would be harder for them to escape outwards. That would explain why
only most massive GCs exhibit variation in Fe content (and other heavy
elements), as described in Sect. \ref{sec:review}.

\subsection{The inward-propagating supershell and the formation of LG stars}
\label{sub:inwss}

Once the supershell has fragmented and formed new stars (the SSSs), we
wish to follow the evolution of the SSS ejecta and analyze under what
conditions can these ejecta, perhaps mixing with some pristine gas,
lead to the formation of new stars. The first problem we face is to
figure out what is the exact location of these SSSs. As we can see
from Fig. \ref{fig:dens_expl}, the supershell is initially pretty
narrow, but it broadens as time goes on. This broadening is partially
deceptive given the logarithmic scale: the large majority of the
supershell mass is confined within $0.5$\,pc behind the shock, even in
the last plotted moment in time. Still, lacking a detailed physical
description of the star formation process, there is still considerable
uncertainty about the birth site of SSSs. One might argue that stars,
being collisionless, are not slowed down by the pressure of the
external medium and so might overtake the supershell. This argument is
however weakened by the fact that stars form out of the densest,
coolest clumps of ISM in the supershell. Pressure gradients between
the center and the supershell are steeper along directions that do not
cross these dense clumps and, as is well known, large pressure
gradients favour a faster expansion of the supershell. Moreover,
supershell expansion in non-uniform media may lead to the development
of Rayleigh-Taylor instabilities. The clumps of denser gas formed as a
consequence of the Rayleigh-Taylor instabilities are more severely
decelerated and tend to lag behind the expanding shock \cite[see
e.g.][]{mfa91, dpj96}. For this reason, we decide to initially place
the SSSs close to the base of the supershell, at a distance of $\sim$
$16$\,pc from the center. We label this {\it Model~1}. This choice has
the further advantage that the spherical volume of radius $16$\,pc has
been almost completely evacuated of pristine gas. Only about 8
$\times$ 10$^3$ M$_\odot$ of gas are contained in this volume.
Therefore, the ejecta of the SSSs propagating inwards are mixed with
just a small fraction of pre-existing gas. As a comparison, we
consider also another {\it Model~2} in which the SSSs are placed close
to the density peak of the supershell, at a distance of $20.6$\,pc
from the center.

\begin{figure}[t]
\includegraphics[width=5.6cm, angle=270]{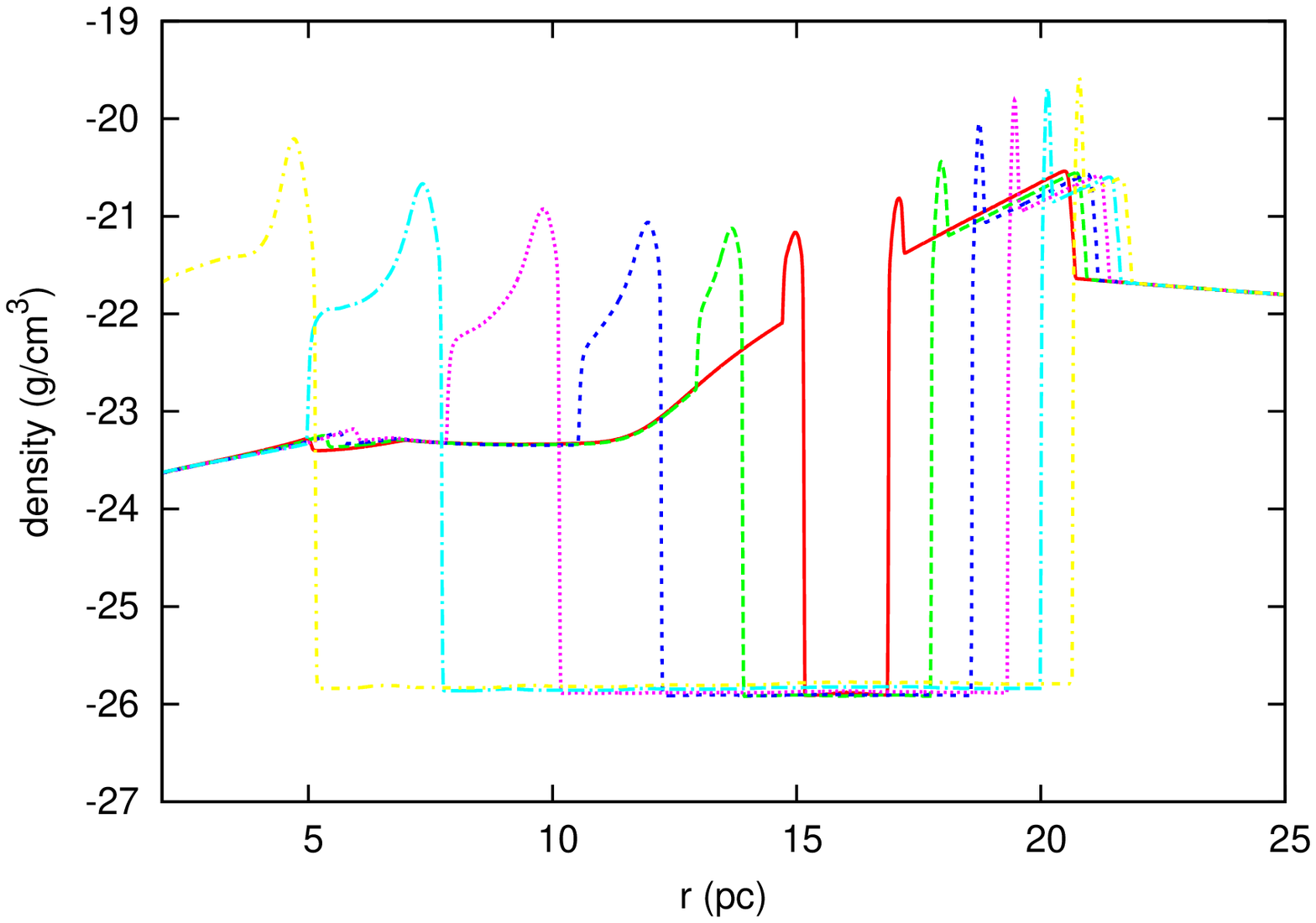}
\caption{Evolution of the gas density (in g cm$^{-3}$) after the
  formation of SSSs for Model 1. The density is plotted every 0.1 Myr
  after $t = t_{frg}$ starting with the solid red line.}
\label{fig:dens_feed}
\end{figure}

As already mentioned, the feedback from the SSSs is calculated by
means of tailored Starburst99 simulations. The evolution of density
after the formation of SSSs for Model1 is shown in Fig.
\ref{fig:dens_feed}. It is clear that the stellar winds propagate in
part towards the center, piling up the little gas remaining in the
cavity. On the other hand, the original supershell continues its
propagation outwards. A two-shell structure is formed. The inner
supershell cools relatively quickly and becomes denser. The
fragmentation time, calculated again as described in Sect.
\ref{subs:fragm}, becomes equal to the evolutionary time at $t\simeq$
0.6 Myr after the insertion of the SSSs, which also the last moment in
time plotted in Fig.  \ref{fig:dens_feed}. Notice that, in order to
calculate the fragmentation time, we use the positive sign of the term
${\alpha^2}/{t^2}$ in Eq. \ref{eq:fragtime}. However, this term turns
out to be much smaller than the ${G^2 \Sigma^2}/{4 a_s^2}$ term. At
this moment, the inner supershell has reached a distance of $\sim$ 5
pc from the center, and has accumulated slightly more than 10$^4$
M$_\odot$ of gas. A significant, but still relatively low (about 20
\%) fraction of this gas is made of ejecta of SSSs, which were able to
cool in a short timescale. This short cooling timescale is due to the
large densities of the supershell. The fact that the fraction of SSS
ejecta in the inward-propagating supershell is relatively low is due
to the fact that, with our adopted Starburst99 model, the amount of
material ejected by stellar winds in the first Myr is quite low. It
increases and becomes significant only after 3-4 Myr.

\begin{figure}[t]
\includegraphics[width=5.6cm, angle=270]{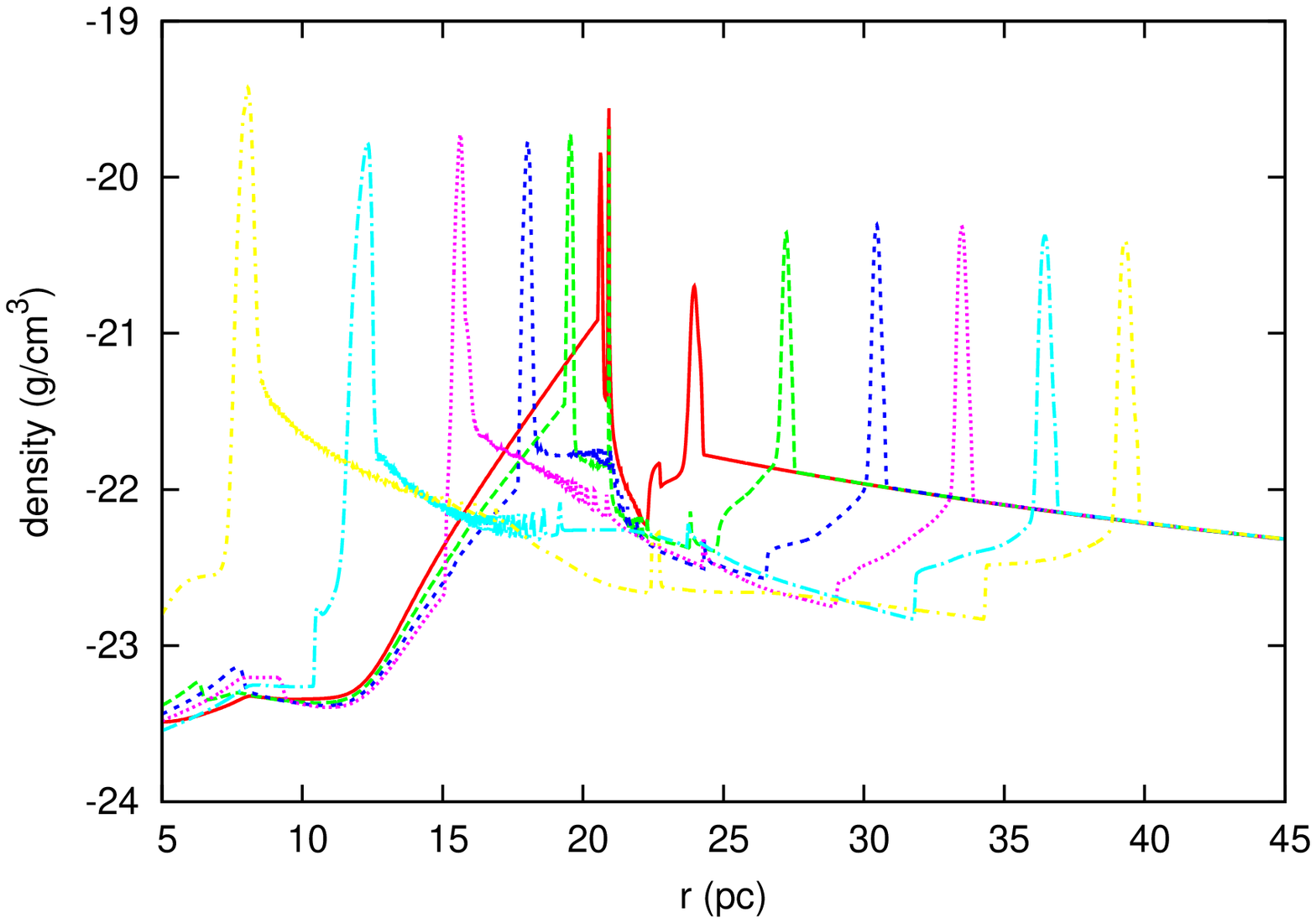}
\caption{Evolution of the gas density (in g cm$^{-3}$) after the
  formation of SSSs for Model 2. The density is plotted every 0.5 Myr
  after $t = t_{frg}$ starting with the solid red line.}
\label{fig:dens_feed_m2}
\end{figure}

{\it Model~2} takes a longer time to propagate towards the center and
to fragment, mainly due to the higher densities the supershell has to
travel through. The fragmentation occurs after $\sim$ 3 Myr, when the
inward-propagating supershell has reached a distance of $\sim$ 8 pc
from the center (see Fig. \ref{fig:dens_feed_m2}). The mass of the
unstable supershell at this point is relatively large (about 10$^5$
M$_\odot$), but the majority of this gas is pristine, thus the ejecta
of the SSSs can not pollute significantly the new generation of stars.
The energy of the SSSs is simply pushing the pre-existing supershell
inwards. Notice that at an age of 3 Myr the first SNe start exploding.
Although we do not see the signs of contamination of SN ejecta in this
specific model, it is at least conceivable that some GCs might form LG
stars partially polluted by ejecta of Type II SNe.

\section{Discussion and conclusions}
\label{sec:disc}

In this work we have studied the explosion of a very energetic Pop-III
star and the subsequent supershell expansion. This supershell becomes
gravitationally unstable and fragments and, eventually, starts forming
stars. Depending on the explosion energy and on the central density of
the ISM surrounding the Pop-III star, gravitational instabilities
start growing $\sim$ 0.5--5 Myr after the Pop-III explosion, and
eventually lead to the formation of supershell stars (SSSs). The fate
of the ejecta of these SSSs is the main aim of this paper. We have
seen that the energy of SSS stellar winds is able in part to
accelerate the already existing supershell created by the Pop-III, in
part to create a new, inward-propagating supershell. With our
assumptions, this supershell is particularly rich in SSS ejecta, at
least if we assume that the SSSs are initially located close to the
inner edge of the supershell.  This supershell is able to cool in a
very short time (a fraction of a Myr), due to its very large density.
We have also checked that this inward-propagating supershell also
becomes gravitationally unstable and might lead to a new generation of
stars.  There is thus the possibility to repeat the already described
process and create a new inward-propagating supershell and, from it, a
new stellar population.  For the moment we have limited our analysis
to the study of the formation of this second generation of stars, but
we intend to analyze the possibility to form a third generation of
stars in a future work.

We have considered also if our depicted scenario can help explaining
some of the puzzling aspects of globular clusters (GCs). In
particular, we wanted to see whether it can explain some
characteristics of GC stars listed in Sect. \ref{sec:review}, namely
specificity, predominance, discreteness and supernova avoidance.

Our scenario certainly helps explaining the specificity of GCs, as it
singles out one physical process which was present only in the early
Universe (Pop-III stars); see also the discussion in Sect.
\ref{sec:idea}. It can help also explain the predominance of late
generation (LG) stars as compared to the first generation. We have
seen that, depending on the explosion energy and on the central
density, the supershell can fragment and form new stars at a distance
between 4 and 20 pc from the center of the explosion. LG stars form
much closer to the cluster center, so they have a higher chance to
remain bound to the cluster than first generation (1G) stars (our
SSSs). Moreover, SSSs are born with a positive radial velocity, at
variance with LG stars, and this increases the chance that a large
fraction of 1G stars might be unbound. Even if this argument is sound,
our simulations can not make quantitative predictions on that and we
have to wait for N-body simulations to shed light on the different
dynamics and on the fate of 1G and LG stars. As explained in Sect.
\ref{sec:review}, \citet{milo17} find a clear correlation between the
mass of a GC and the fraction of 1G stars. This fraction is large for
low-mass clusters and decreases with increasing GC mass. Also this
result is broadly consistent with our scenario. In fact, we expect
that, if the supershell formed by the Pop-III star is large, then a
large number of 1G stars can be formed. These stars however, due to
the large distance of the supershell, have a larger probability of
escaping the gravitational potential. As we have seen (Sect.
\ref{sec:idea}), the dependence of the chemical complexity in GCs on
the stellar mass might also be accounted for in our model.  However,
it was not our intention in this paper to provide a detailed model of
the chemical composition of GCs. The abundances of different elements
are taken from a Starburst99 model, and this is probably inadequate to
account for the chemical peculiarities of GCs. The time evolution of
the content of He, C, N, O, Na, Si and of other elements will be the
subject of a subsequent paper.  Also as regards the dynamics of stars,
although the argument presented here is sound, still we have to wait
for N-body simulations for a detailed quantitative analysis.

We conclude this paper with a quick discussion on the comparison
between GCs in the Milky Way and in the Magellanic Clouds.  Most GCs
in the LMC and SMC with ages smaller than $\sim$2 Gyr show an extended
main-sequence turn off \citep{mack08, milo09, milo16}. Although the
connection between this spread and the presence of multiple stellar
populations is still not firmly established, the impression is that
young and intermediate-age clusters in Magellanic clouds are the
counterparts of old GCs with multiple populations \citep{cs11}. This
connection is fascinating although only measurements of abundances of
individual stars (for which we must wait JWST) can tell us whether we
might face for GCs in the Magellanic Clouds the same problems we face
for Milky Way GCs.  For the moment we note that the model of a
collapsing supershell may also apply to intermediate age clusters in
the LMC and SMC if there was a sufficiently strong source of energy.
This source of energy might also be due to a Pop-III star, but this is
unlikely to be the case for young GCs.

\acknowledgments

Support for this project was provided by the Czech Science Foundation
grant 15-06012S and by the project RVO: 6785815. We thank the
anonymous referee for very useful and insightful remarks and
suggestions.  We thank Sona Ehlerova and Anthony Whitworth for
assistance with the preparation of the manuscript.

\nocite{*}
\bibliography{myref}

\appendix

\section{The role of the external pressure on supershell fragmentation}

\label{apppressure}

\citet{franta17} suggest that the relative importance of the pressure
$P_{ext}$ confining a shell influences the fragmentation not only
quantitatively, but also qualitatively.  If the confining pressure
dominates the self--gravity, the shell breaks into many
gravitationally stable objects which subsequently coalesce until they
form enough mass to collapse gravitationally.  If the self--gravity
dominates the confining pressure, the condensing fragments collapse
directly into gravitationally bound objects.  While the dispersion
relation proposed by \citet{ee78} provides a reasonable estimate for
fragment properties for self--gravity dominated shells, no linearised
estimate holds for pressure dominated shells.  For pressure dominated
shells, \citet{franta17} indicate that the Jeans mass is a proxy for
the fragment mass.  This determines our choice of fragment radius $d$
and the approximations made in \eq{efrag} in these two cases.  For
pressure--dominated shells, we assume that the most unstable fragment
contains one Jeans mass while for self--gravity dominated shells, we
assume that the radius corresponds to the most unstable wavelength in
the dispersion relation proposed by \citet{ee78}.  Therefore,
\begin{subnumcases}{d =}
\frac{\sqrt{1.5 A} a_s^2}{G \Sigma} & pressure dominated shell \label{eradiusfragpress} \\
\frac{2 a_s^2}{G \Sigma}. & self-gravity dominated shell \label{eradiusfragsg}
\end{subnumcases}
where parameter
\begin{equation}
A = \frac{1}{\sqrt{1 + 2P_{ext}/(\pi G \Sigma^2)}}
\label{epara}
\end{equation}
indicates the relative importance of the external pressure versus
self--gravity \citep{ee78}.  Parameter $A$ lies in the interval $(0,
1)$, where pressure dominated shells have $A$ near zero and
self--gravity dominated ones $A$ near unity.

Since fragmentation occurs at substantially different times for
pressure dominated and self--gravity dominated shells, we investigate
which case is more appropriate to our supershell model and use the
approximation for the fragmenting time accordingly.  Let we assume
that the supershell fragments in the pressure dominated case.  For
this reason, we omit the second term on the right hand side of
\eq{efrag} because unstable fragments are delivered by coalescence
with no role of the pressure gradient.  The sound speed of the
supershell $a_s$ is insensitive to the position within the supershell
wall (see Figure \ref{fig:ss}).  The sound speed also changes very
little during the expansion.  This enables us to express the
fragmenting time in the closed form.

Let us assume a supernova explosion at the density peak of a medium
with a general power--law density distribution $n(r) = n_c
(r/r_c)^{-\omega}$. The exponent of the expanding law $\alpha$
(equation \ref{eexpansionlaw}) depends on the value of $\omega$ and
whether the shell is in the energy conserving phase or the momentum
conserving phase.  While for expansion in the energy conserving phase
$\alpha = 2/(5 - \omega)$, for expansion in the momentum conserving
phase $\alpha = 1/(4 - \omega)$ \citep{omk88}.

The surface density of a supershell of radius $R$ is 
\begin{equation}
\Sigma(R) = \frac{M(R)}{4 \pi R^2} = \frac{n_c \bar{m} r_c^2 K^{1-\omega} t^{\alpha(1-\omega)}}{3 - \omega},
\label{esigma}
\end{equation}
where $\bar{m}$ denotes the mean gas particle mass.  The supershell is
confined by two pressures: thermal pressure from the cavity acting on
its inner surface, and ram pressure from the accreting medium acting
on its outer surface.  In the momentum conserving phase, the latter of
the pressures is significantly higher than the former.  For the
adopted model of shell fragmentation, it is not a priori clear, which
of these pressures characterises fragmentation better.  We choose the
ram pressure, which is higher, to represent $P_{ext}$, and we shall
see that the shell fragments in self--gravity dominated case, so the
choice of the particular pressure is not important.  Thus,
\begin{equation}
P_{ext} = n(r) \bar{m} \left( \nderrow{R}{t} \right)^2 = n_c \bar{m} r_c^2 K^{2 - \omega} \alpha^2 t^{2\alpha - \alpha \omega -2}.
\label{epext}
\end{equation}
Substituting \eq{esigma}, \eq{epext} and the parameter $A$ for
pressure dominated shells (small $A$; $P_{ext} \gg \pi G \Sigma^2$)
where \eq{epara} becomes
\begin{equation}
A \simeq \left\{ \frac{\pi G \Sigma^2}{2 P_{ext}} \right\}^{1/2} = \sqrt{\frac{\pi G n_c \bar{m} r_c^2}{2}} 
\frac{K^{-\omega/2} t^{-(\alpha \omega + 2)/2}}{\alpha (3 - \omega)},
\label{eparaaprox}
\end{equation}
into \eq{eradiusfragpress} and \eq{efrag}, one obtains the fragmenting
time ($t_{frg} = t_{col}$),
\begin{equation}
t_{frg}  = \left\{ \frac{9 \pi a_s^4 (3 - \omega)^6 (1 + \alpha^2)^4 K^{7 \omega}}
{8 \alpha^2 (G n_c \bar{m} r_c^2)^{7} K^8} \right\}^{1/(6 + 8\alpha - 7\alpha \omega)}.
\label{efragpressure}
\end{equation}

We use $\alpha = 1/2$ and $\omega = 2$ for the presented model.
Substituting these values to \eq{efragpressure} and using \eq{epara}
with $t = t_{frg}$, we realise that the supershell fragmented when $A
\simeq 0.8$, i.e. in the self--gravity dominated case. This finding
contradicts the original assumption that the supershell fragments in
the pressure dominated case.

Therefore, to estimate the fragmenting time, we use the estimate of
fragment radius in the self--gravity dominated case and also take into
account the term in \eq{efrag} representing the pressure gradient.
Substituting \eq{eradiusfragsg} into \eq{efrag}, one finds that the
fragment collapse time-scale is given by
\begin{equation}
t_{col} = \left(\frac{G^2 \Sigma^2}{4 a_s^2}-\frac{\alpha^2}{t^2}\right)^{-1/2}.
\label{eq:afragtime}
\end{equation}
Note that for shells expanding in the momentum conserving phase, the
second term on the right hand side of \eq{eq:afragtime} decreases
faster than the first term if $\omega < 5/2$.  Thus, for the choice
$\omega = 2$, there always exists an expansion time $t$ which is
larger than the fragment collapse time $t_{col}$, and the shell
fragments.

\end{document}